\documentclass[aps,prc,twocolumn]{revtex4}
\usepackage{graphicx}
\begin{document}
\title{Action potential as a pressure pulse propagating in the axoplasm}

\author{M.~M.~Rvachev}
\email{rvachev@alum.mit.edu}
%\affiliation{Physics Department, Massachusetts Institute of Technology, Cambridge, Massachusetts 02139, USA}

\date{May 14, 2009}

\begin{abstract}
We suggest that the propagation of the action potential is driven by a pressure pulse propagating in the axoplasm along the axon length. The pressure pulse mechanically activates Na$^+$ ion channels embedded in the axon membrane. This activation initiates the development of a local membrane voltage spike, as in the Hodgkin-Huxley model of the action potential. Extracellular Ca$^{2+}$ ions influxing during the voltage spike trigger a mechanism that amplifies the pressure pulse and therefore prevents its viscous decay. The model is able to explain a number of phenomena that are unexplained within the Hodgkin-Huxley framework: the Meyer-Overton rule for the effectiveness of anesthetics, as well as various mechanical, optical and thermodynamic phenomena accompanying the action potential. The model correctly predicts the velocity of propagation of the nerve impulse, including its dependence on axon diameter, degree of myelination and temperature.
\end{abstract}
% insert suggested PACS numbers in braces on next line
\pacs{}
% insert suggested keywords - APS authors don't need to do this
%\keywords{}

%\maketitle must follow title, authors, abstract, \pacs, and \keywords
\maketitle

\section{Introduction}

The well-known Hodgkin-Huxley theory of propagation of the nerve impulse in neuronal axons \cite{hodg52} has recently been the subject of constructive criticism \cite{rva03,rva03b,heim05,heim06}. As a purely electrical theory, the Hodgkin-Huxley theory is unable to explain a number of mechanical and thermodynamic phenomena that are observed synchronously with a propagating action potential. These include changes in nerve dimensions and in the normal force exerted by the nerve \cite{iwas80b,iwas80,tasa80,tasa82,tasa90}, reversible changes in temperature and heat \cite{abbo58,howa68,ritc85,tasa89,tasa92} and changes in fluorescence intensity and anisotropy of lipid membrane markers \cite{tasa68,tasa69}.  Importantly, within the electrical framework, it has not been possible to explain    
the famous Meyer-Overton rule \cite{meye99,over01} for the effectiveness of anesthetics on nerve fibers. The rule states that the effectiveness of an anesthetic is linearly related to the solubility of that anesthetic in membranes \cite{urba06}. The rule is valid over 5 orders of magnitude and holds independently of the chemical identity of the anesthetic; it is valid for noble gases such as argon and xenon as well as for alkanols. No link is found between the reactiveness of anesthetics with ion channels, the major actors in the Hodgkin-Huxley theory, and the Meyer-Overton rule.

The Hodgkin-Huxley mechanism of propagation of the nerve impulse \cite{hodg52} can be divided into  two phases. First, after an initial depolarization of a membrane segment to the excitation level, a local voltage spike develops. Second, ionic currents flowing through the excited membrane depolarize adjacent unexcited membrane segments to the excitation voltage, causing propagation of the nerve impulse. 
The viability of the first phase of the mechanism is little doubted. 
In the voltage clamp method \cite{hodg52}, electrodes are positioned inside and outside of an axon,   
allowing the experimenter to change the membrane voltage arbitrarily and to measure the resultant membrane voltage and current. Using this method, it is also possible to simultaneously depolarize the membrane along the entire length of the axon. If depolarization reaches a certain threshold level, one observes a "membrane" action potential (voltage spike), which develops concurrently along the axon length. 
Hodgkin and Huxley studied this type of action potential in detail \cite{hodg52}. They showed that the duration, magnitude and time course of the voltage spike are explained by ionic fluxes across the membrane. The fluxes arise due to time- and voltage-dependent changes in ion-specific membrane permeability. There are no axial currents during the membrane action potential, which simplifies theoretical modeling of the process.

The second phase of the Hodgkin-Huxley mechanism, depolarization of unexcited membrane segments by ionic currents straggling in the axial direction along the axon length and thus causing impulse propagation, has not been shown experimentally. To test the validity of this part of the mechanism, one would ideally wish to block axial ionic currents at an axonal cross-section and to observe whether a propagated action potential is able to cross this boundary. To date, such an experiment has not been done. Proof of the propagation mechanism has been based on numerical calculations. Using their electrical framework, Hodgkin and Huxley calculated \cite{hodg52} the velocity of the action potential for the squid giant axon. Although the calculated velocity agreed well with experiment, a number of flaws in the formulation of the problem were pointed out \cite{heim06}; e.g., changes in the membrane capacitance due to variation in membrane thickness were ignored. Hodgkin and Huxley themselves noted \cite{hodg52} that due to non-zero axial currents "the situation is more complicated in a propagated action potential" compared to a membrane action potential, and "it is not practicable to solve as it stands" the equation describing the problem. To simplify, they assumed that a steadily propagating solution exists, in which, additionally, the shape of the voltage spike is preserved in time. No analysis of the stability of the solution was made. Even with these simplifications, the numerical calculation was in parts "excessively tedious", and "the solution was continued as a membrane action potential".  Clearly, the calculation was rather complicated and will be expected to have associated error.

In this paper, we present an alternative theory for the propagation of the action potential. We suggest that its propagation is driven by a pressure pulse propagating in the axoplasm. This model allows a well-understood calculation of the action potential velocity. For both myelinated and unmyelinated fibers of various diameters, the velocity predicted by this theory agrees very well with experiment. The theory is able to quantitatively explain the Meyer-Overton rule as well as various mechanical, optical and thermodynamic phenomena that accompany the action potential.
The speculation that density pulses may be related to nerve impulses has been discussed by various authors since 1912 \cite{wilk12,wilk12b,hodg45,kauf89a,kauf89b}.
Most recently, it has been suggested \cite{heim05,heim06} that the action potential is a soliton propagating in the axon membrane.  While this model appears to account for the abovementioned phenomena, as well as for the Meyer-Overton rule, it does not seem to explain changes in the propagation velocity that are related to axon diameter \cite{hill77}.

\section{The Hodgkin-Huxley theory}
According to the Hodgkin-Huxley theory \cite{hodg52}, the propagation of the action potential is driven by electrical currents flowing in the axial direction along the length of the axon. In the resting state, ion transporters maintain transmembrane concentration gradients, pumping K$^+$ inside the cell and Na$^+$ outside \cite{hill77}. The resting membrane is more permeable to K$^+$; therefore, the electrochemical equilibrium maintains a negative potential inside the cell. Depolarization of the membrane to a threshold level results in a rapid and transient opening of voltage-gated Na$^+$ channels, causing the membrane voltage to spike to the Nernst equilibrium for Na$^+$ ions. 
This is followed by a slower activation of voltage-gated K$^+$ channels, which repolarizes the membrane to the resting potential.
The action potential propagates due to axial spread of electrical currents from the site of the voltage spike \cite{hodg52}. These currents depolarize adjacent unexcited regions of the membrane to the excitation voltage.
Within this framework, the membrane is described as the electrical circuit shown in Fig.~1. 
\begin{figure}[t]
  \begin{center}
%    \showthe\columnwidth % Use this to determine the width of the figure.
    \includegraphics[width=0.9\columnwidth]{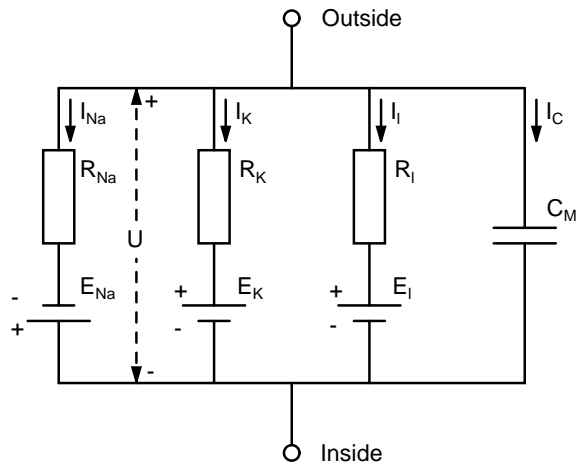}
    \caption{\label{fig:mem_cir} Electrical circuit representing membrane. $R_{Na}=1/g_{Na}$, $R_{K}=1/g_{K}$, $R_{l}=1/g_{l}$. $R_{Na}$ and $R_{K}$ vary with time and membrane potential, the other
    components are constant. From \cite{hodg52}.}
  \end{center}
\end{figure}
The current density across the membrane is given by:
\begin{eqnarray}
I & = & C_M\frac{dU}{dt}+g_K(U-E_K)+ \\
& & g_{Na}(U-E_{Na})+g_l(U-E_l), \nonumber
\end{eqnarray}
where the four terms on the right-hand side of the equation give, respectively, the membrane capacity current, the current carried by K$^+$ ions, the current carried by Na$^+$ ions, and the "leakage current" (due to Cl$^-$ and other ions) per unit area of membrane. $C_M$ is the membrane capacitance, and $g_K$, $g_{Na}$ and $g_l$ are, respectively, the conductances to K$^+$ ions, Na$^+$ ions and the leakage currents. 
$E_K$,  $E_{Na}$ and  $E_l$ are, respectively, the equilibrium potentials for K$^+$ and Na$^+$ and the leakage currents.  $g_{Na}$ and $g_K$ vary with the membrane potential $U$ and time as parameterized from voltage clamp data \cite{hodg52}, while $C_M$, $g_l$, $E_K$,  $E_{Na}$ and $E_l$ are constant.
For a "membrane" action potential, $U$ is the same at each instant over the length of the axon. There is no current along the axon axis, and $I$ is always zero. The time course of the action potential is obtained by solving eq. (1) numerically, with $I=0$ and the initial depolarization condition $U=U_0$. Hodgkin and Huxley showed convincingly \cite{hodg52} that the time course of the membrane action potential is indeed explained by behavior in $g_{Na}$ and $g_K$, as parameterized from the voltage clamp data, and the dynamics as described by eq. (1).

For reconstructing a propagated action potential, one has to consider local circuit currents, which leads to \cite{hodg52}: 
\begin{equation}
I = \frac{R}{2\rho_2}\frac{\partial^2U}{\partial x^2},
\end{equation}
where $R$ is the axon radius, $\rho_2$ is the specific electrical resistance of the axoplasm, and $x$ is distance along the axon.
Combining eqs. (1) and (2), one obtains:
\begin{eqnarray}
\frac{R} {2\rho_2} \frac{\partial^2U}{\partial x^2}&=&C_M \frac{\partial U}{\partial t}+g_K(U-E_K)+\\
& & g_{Na}(U-E_{Na})+g_l(U-E_l). \nonumber
\end{eqnarray}
Hodgkin and Huxley noted that this equation was not practicable to solve in this form. To simplify, they assumed that a steadily propagating solution exists, in which the shape of $U$ against distance does not change in time, leading to:
\begin{equation}
\frac{\partial^2 U} {\partial x^2} = \frac{1}{\theta^2} \frac{\partial^2 U} {\partial t^2},
\end{equation}
where $\theta$ is the velocity of conduction. From eqs. (3) and (4), one obtains:
\begin{eqnarray}
\frac{R} {2\rho_2\theta^2} \frac{d^2U}{dt^2} & = & C_M \frac{dU}{dt}+g_K(U-E_K)+ \\
& & g_{Na}(U-E_{Na})+g_l(U-E_l). \nonumber
\end{eqnarray}
This is an ordinary differential equation and can be solved numerically if $\theta$ is known.
It is solved by guessing a value of $\theta$, inserting it in eq. (5) and carrying out numerical integration starting from the resting state at the foot of the action potential. As the guessed $\theta$ is too small or large, it is found that $U$ approaches either $+\infty$ or $-\infty$. A new value of $\theta$ is then chosen and the procedure repeated until a correctly guessed $\theta$ brings $U$ back to the resting condition when the action potential is over \cite{hodg52}. Although the Hodgkin-Huxley model does lead to a propagating action potential, certain flaws were found in the formulation of the problem \cite{heim06}. Importantly, the model is not able to explain a number of mechanical, optical and thermodynamic phenomena accompanying the action potential, as well as it provides few clues for understanding the Meyer-Overton rule \cite{heim05,heim06}.
\section{The pressure wave model}
We suggest that the propagation of the nerve impulse is driven by a pressure pulse propagating in the axoplasm along the axon length. While several variations of the process can be envisioned, we suggest that it proceeds as shown in Fig. 2. 
\begin{figure*}[t]
  \begin{center}
%    \showthe\twocolumnwidth % Use this to determine the width of the figure.
    \includegraphics[width=1.5\columnwidth]{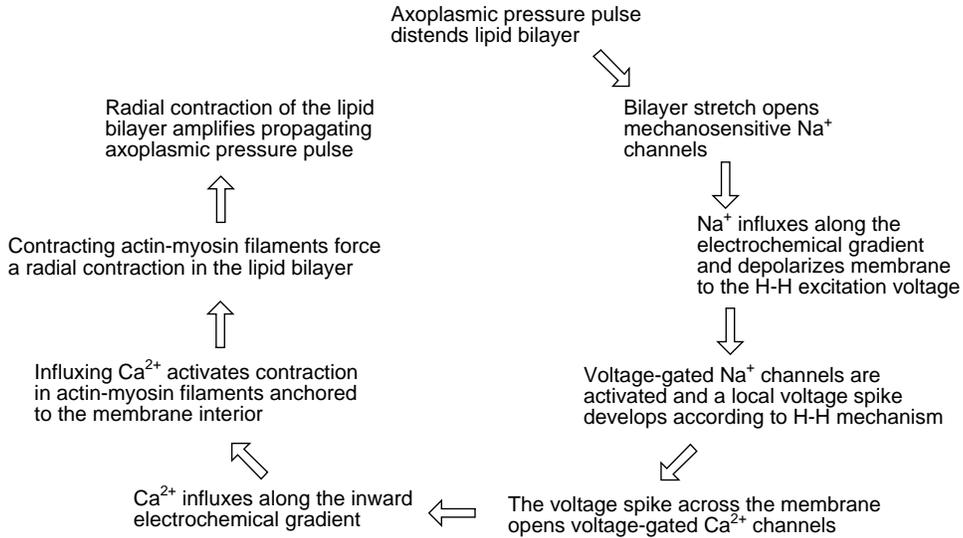}
    \caption{\label{fig2} Diagram of the suggested process of propagation of the axoplasmic pressure pulse and the action potential.}
  \end{center}
\end{figure*}
A propagating axoplasmic pressure pulse distends the axon membrane, which causes mechanical activation of the membrane Na$^+$ ion channels. Activated Na$^+$ channels allow Na$^+$ ions to influx along the inward electrochemical gradient, thus locally depolarizing the membrane to the excitation voltage and initiating a local Hodgkin-Huxley voltage spike. The local transmembrane voltage spike activates membrane voltage-gated Ca$^{2+}$ channels, leading to an influx of Ca$^{2+}$ ions along their inward electrochemical gradient. The presence of free intracellular Ca$^{2+}$ induces a contraction in actin-myosin filaments anchored to the membrane interior. The filament contraction forces a radial contraction in cylindrical segments of the membrane (Fig. 3),
\begin{figure}[t]
  \begin{center}
%    \showthe\columnwidth % Use this to determine the width of the figure.
    \includegraphics[width=\columnwidth]{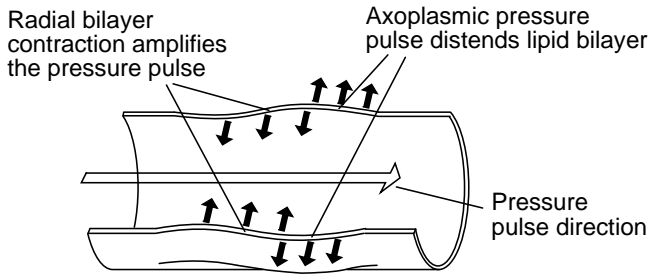}
    \caption{\label{fig:fig3} Amplification of the axoplasmic pressure pulse by a radial contraction  of the axon membrane (drawn not to scale).}
  \end{center}
\end{figure}
which amplifies the propagating pressure pulse and thus compensates for its decay due to viscosity. (In section 5, it is shown that under physiological conditions axoplasmic pressure pulses decay over roughly 1 mm distances.)

Figs. 4(a)-(b)
\begin{figure}[t]
  \begin{center}
%    \showthe\columnwidth % Use this to determine the width of the figure.
    \includegraphics[width=0.85\columnwidth]{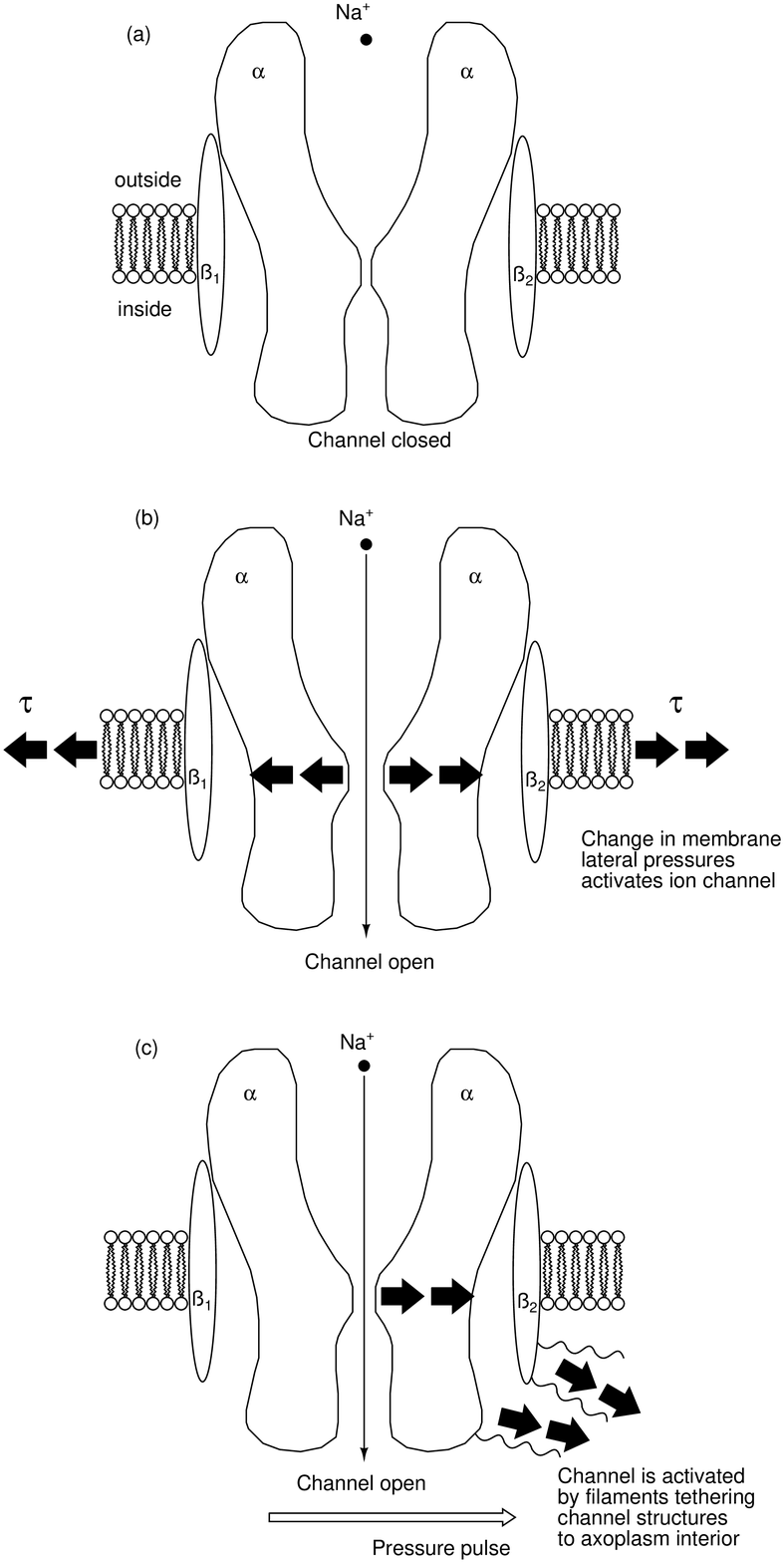}
    \caption{\label{fig4} Mechanical activation of Na$^+$ ion channels by the traveling pressure pulse (see text).}
  \end{center}
\end{figure}
illustrate a suggested process of  mechanical activation of Na$^+$ ion channels. The lipid bilayer is stretched laterally as the propagating axoplasmic pressure pulse causes a $\Delta R(t)$ increase in the membrane cylinder radius $R$, corresponding to a $\Delta\tau(t)$ increase in the lipid bilayer tension $\tau$. Tension $\tau$ can be expressed through the bilayer lateral pressure profile $p(z)$ \cite{cant01}: $\tau=\int_{-\frac{h}{2}}^{\frac{h}{2}}p(z) dz$, where $z$ is the depth within the bilayer, and the integral is over the bilayer thickness $h$. The change $\Delta\tau(t)$ in the bilayer tension results in a change $\Delta p(z,t)$ in the lateral pressure profile such that $\Delta\tau(t)=\int_{-\frac{h}{2}}^{\frac{h}{2}}\Delta p(z,t) dz$. We suggest that these alterations in the bilayer lateral pressures modulate conformational opening of Na$^+$ channels imbedded in the bilayer. Alternatively, it is plausible that Na$^+$ channels are activated by direct mechanical links (such as filaments) tethering the ion channel structures to the interior cytoskeleton that is perturbed by the propagating pressure pulse, as shown in Fig. 4(c).

According to the model presented, an initial influx of Na$^+$ ions through mechanically activated ion channels depolarizes the membrane to the excitation voltage, leading to the development of a local Hodgkin-Huxley voltage spike. Further, the change in voltage across the membrane activates membrane voltage-gated Ca$^{2+}$ channels, allowing entry of Ca$^{2+}$ ions into the cell, which initiates a contraction in the filament network anchored to the membrane interior. Ca$^{2+}$ is a good candidate for mediating the Hodgkin-Huxley voltage spike into the filament contraction for several reasons. First, 
Ca$^{2+}$ ion channels are found in virtually every excitable cell 
\cite{shep94}, and many types of Ca$^{2+}$ ion channels are directly and rapidly gated by voltage \cite{nau08}. Furthermore, an increase in free intracellular Ca$^{2+}$ is often associated with initiation of motion in cells, from motility in freely moving cells and muscle contraction to synaptic vesicle release at synapses \cite{shep94}. Also, the free intracellular Ca$^{2+}$ concentration is typically extremely low, while it is much larger outside the cell.
Ca$^{2+}$ ions that influx during the action potential 
should instantly increase manyfold the free Ca$^{2+}$ concentration in the vicinity of the filament
network attached to the inner membrane surface; that, coupled with the ability of Ca$^{2+}$ to quickly
induce conformational changes in proteins (such as upon its binding
to actin filaments in the actin-myosin muscle complex), could provide a
fast contractile response necessary to amplify a propagating pressure
pulse. In this scheme, the Hodgkin-Huxley sodium and potassium currents and the voltage spike simply provide a means for activating voltage-gated Ca$^{2+}$ ion channels. Alternatively, it is possible that amplification of the pressure pulse proceeds through an electromechanical coupling mechanism, such as voltage-induced membrane movement \cite{zhan01} resulting from the Hodgkin-Huxley voltage spike. Note that, since the radial bilayer contraction follows the pressure pulse at the same velocity, even a small contraction yielding a small amplification will accumulate over distance and will create a larger amplifying effect. 
In this context, it should be noted that a pressure pulse propagating in an inviscid incompressible fluid enclosed in a viscoelastic tube, dissipates energy and decays because the pressure exerted by the tube on the fluid during tube radial expansion is greater than that during tube radial contraction; therefore, the total work done by the tube on the fluid is negative. In axons, the situation may be the reverse -- a larger pressure exerted by the bilayer during its contraction may result in an overall transmission of kinetic energy to the axoplasm.

\section{Meyer-Overton rule and the action of anesthetics}
Within the framework of the nerve impulse as an axoplasmic pressure pulse, we naturally arrive at a quantitative explanation of the Meyer-Overton rule for inhaled anesthetics \cite{meye99,over01,urba06}. Essentially, the rule states that anesthetic potency is determined by the bilayer concentration of anesthetic, independent of its molecular identity \cite{cant01}. Therefore, the Meyer-Overton rule can be explained if there is a property of the lipid bilayer that is essential for nerve impulse propagation and that, in addition, depends only on the bilayer concentration of anesthetic. As is shown below, one such property may be the bilayer area expansion modulus, which affects both the decay and the velocity of axoplasmic pressure pulses. It is also shown that anesthetics may inhibit the mechanical activation of membrane Na$^+$ channels by the traveling pressure pulse (and therefore suppress the conduction of the action potential) in a manner consistent with the Meyer-Overton rule.

Suppose that an anesthetic molecule is dissolved within the lipid bilayer (anywhere from the headgroup to the hydrophobic region), and it weakens and destabilizes the bilayer structure nonspecifically, e.g., through creating a
defect in the packing of lipid molecules and thereby nonspecifically disordering lipid chains (Fig. 5(a)). 
\begin{figure}[t]
  \begin{center}
%    \showthe\columnwidth % Use this to determine the width of the figure.
    \includegraphics[width=0.8\columnwidth]{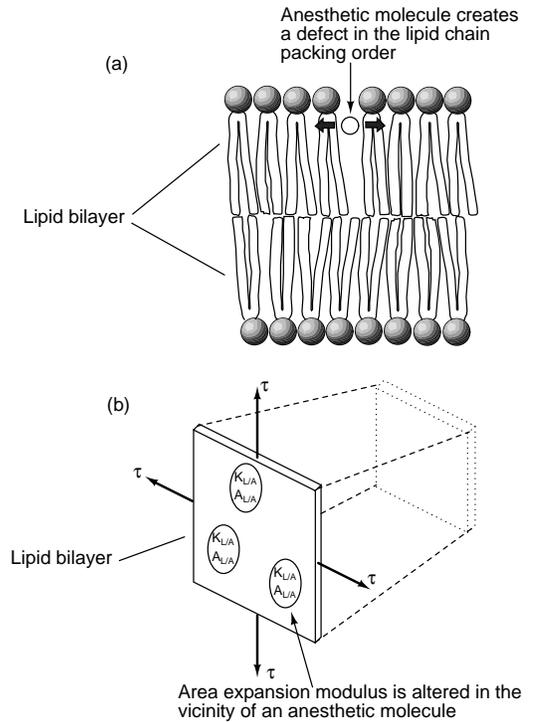}
    \caption{\label{fig5} (a) Disordering of lipid packing order by an anesthetic molecule dissolved in the bilayer. (b) Pockets of altered area expansion modulus in the vicinity of anesthetic molecules.}
  \end{center}
\end{figure}
Such disordering interaction will generally decrease the bilayer elastic area expansion modulus in the vicinity of the lipid/anesthetic molecular complex (it is well established that bilayers with larger lipid chain disordering generally display lower elastic moduli and tensile strengths \cite{need95}). Using a simple macroscopic model of the membrane  \cite{need95}, shown in Fig. 5(b), the total membrane area expansion modulus can be expressed as:
\begin{equation}
K=\left(\frac{a_M}{K_M}+\frac{a_{L/A}}{K_{L/A}}\right)^{-1},
\end{equation}
where the right-hand side of the equation is a combination of the area expansion moduli of the two components, namely, non-anesthetized membrane $K_M$ (which describes the elasticity of the lipid bilayer, embedded proteins and other native structures) and a purported lipid/anesthetic molecular complex $K_{L/A}$ ($K_{L/A}<K_M$), scaled by their area fractions $a_M$ and $a_{L/A}$. Assuming that anesthetic molecules are dissolved only in the lipid segments of the membrane and that they do not cluster together, the area fractions $a_M$ and $a_{L/A}$ are given by: 
\begin{equation}
a_{L/A} = 1 - a_M = n_{A} \cdot h \cdot A_{L/A},
\end{equation}
where $n_{A}$ represents the concentration of anesthetic molecules within the membrane, in units of molecules per volume, $h$ the bilayer thickness, and $A_{L/A}$ the membrane area occupied by a single lipid/anesthetic molecular complex. Note that $A_{L/A}$ and $K_{L/A}$ are independent of the chemical identity of the anesthetic, due to the purported nonspecificity of anesthetic/lipid interaction. Combining eqs. (6) and (7):
\begin{equation}
K=\left(\frac{1- n_{A} h  A_{L/A}}{K_M}+\frac{ n_{A}  h  A_{L/A}}{K_{L/A}}\right)^{-1},
\end{equation}
where the right-hand side of the equation does not depend on the identity of the anesthetic, but only on its bilayer concentration $n_A$. As will be shown in the next section, both the velocity and the decay length of axoplasmic pressure pulses vary as $\sqrt{K}$. Therefore, a decrease in the area expansion modulus $K$ due to an increase in the anesthetic concentration $n_A$ will reduce the pressure pulse velocity and increase its decay, which, when $n_A$ is of sufficient magnitude, should lead to an inhibition of the pulse propagation. From the form of eq. (8), it follows that anesthetics acting in such a way will comply with the Meyer-Overton rule. From the line of argument above, it is clear that any nonspecific anesthetic/lipid interaction that results in decreasing $K$ with increasing $n_A$ and thus leads to a block in nerve impulse conduction will be consistent with the Meyer-Overton rule.

Another possible mechanism of action of anesthetics consistent with the Meyer-Overton rule relies on inhibition of mechanoactivation of Na$^+$ ion channels by the traveling pressure pulse, in cases where such activation is modulated by changes in the membrane lateral pressures (Fig. 4(b)). As before, we assume that anesthetic molecules dissolved in the bilayer weaken it nonspecifically and that increasing anesthetic concentration $n_A$ decreases the membrane area expansion modulus $K$. As will be discussed in the next section, the potential energy of a propagating axoplasmic pressure pulse is mainly stored in the membrane strain. The membrane strain energy, $E_p$, is proportional to $L\Delta\tau^2/K$, where $L$ is the pulse length and $\Delta\tau$ is the peak increase in the membrane tension $\tau$ from equilibrium. Since $L\sim\sqrt{K}$ (assuming a fixed pulse duration, see eq. (14)), for the same pulse energy ($E_p$ fixed), it follows that $\Delta\tau\sim K^{1/4}$. Therefore, a decrease in the area expansion modulus $K$ will result in a reduced tension increase $\Delta\tau$. At sufficiently high $n_A$ (low $K$), the decreased $\Delta\tau$ may not suffice for activation of mechanosensitive Na$^+$ channels, suppressing the process of pulse amplification and leading to a block of nerve impulse conduction. It should also be noted that, if the anesthetic/lipid interaction changes the lipid bilayer area without affecting its area expansion modulus, the equilibrium tension in the membrane may change, which may affect gating of mechanosensitive Na$^+$ channels, possibly leading to conduction block. This would also conform to the Meyer-Overton rule if the anesthetic/lipid interaction is nonspecific.
In this context, it should be noted that an anesthetic/lipid interaction altering a mechanical property of the lipid bilayer in a nonspecific manner would probably involve a defect-like structure, in which physical properties of the anesthetic compound are not important provided that the compound satisfies certain criteria for creating the defect (e.g., being hydrophobic, nonpolar, and nonreactive and possessing certain spatial and mass dimensions).

\section{Propagation of axoplasmic pressure pulses}
We consider the propagation of small amplitude, axially symmetric pressure pulses in a viscous compressible axoplasm enclosed in a circular cylindrical thin-walled distensible membrane. The theoretical analysis of waves in viscous fluids enclosed in tubes has been presented in \cite{rayl45,morg54,rubi78}. For the limiting case of a viscous compressible axoplasm in a rigid tube, the pressure pulse is an axoplasmic density disturbance that propagates along the axon similarly to a sound wave packet, with the pulse potential energy stored in the axoplasm bulk deformation \cite{rayl45,rubi78}. In the opposing limit of a viscous incompressible axoplasm in a distensible membrane, the pressure pulse manifests itself as an increase in the axon diameter and  associated membrane area expansion, with the pulse potential energy stored in the membrane strain \cite{morg54,rubi78}.
Let us consider a pressure pulse of central frequency $\omega$ propagating in the axoplasm of density $\rho$, compressibility $\kappa$ and dynamic viscosity $\mu$, enclosed in a cylindrical membrane of radius $R$, thickness $h$ and area expansion modulus $K$. Let us also assume that for the deformation of area expansion (when the membrane material behaves in an elastic solid manner), the membrane Young's modulus is $E$ and the Poisson's ratio is $\nu$.
If viscosity is neglected, the pulse propagates with the velocity \cite{chev93}:
\begin{equation}
v_0=\sqrt{\frac{1}{\rho(\kappa + \frac{2R}{K})}}.
\end{equation}
For typical unmyelinated axons, i.e., assuming $K=0.8$~N/m (twice the single plasma membrane area expansion modulus of 0.4 N/m \cite{thom97}, to account for the combination of the axon and adjoining glial cell membranes), 
$\kappa =4.04\cdot10^{-10}$ Pa$^{-1}$ (water at 38 $^\circ$C \cite{lide02}), $\rho = 1050$~kg/m$^3$ (axoplasm \cite{keyn51}), and $2R=1$ $\mu$m, eq. (9) yields the pressure pulse velocity of $28$ m/s. 
Note that if the axon membrane is assumed to be indistensible ($\frac{2R}{K}\ll\kappa$), eq. (9) yields a pulse velocity of 1535 m/s. 
\subsection{Viscous axoplasm}
Viscous forces in the axoplasm and in the axon membrane in general decrease the velocity of axoplasmic pressure waves and introduce a decay \cite{rayl45,morg54,rubi78}. The membrane viscous tension is characterized by the ratio of the surface viscosity (on the order of 10$^{-6}$ N$\cdot$s/m \cite{evan76}) and the time scale of the pressure pulse (here assumed to be close to the action potential duration, about 1 ms), yielding 10$^{-3}$ N/m. Since this value is much smaller than the membrane elastic area expansion modulus, 0.4 N/m, the effects of membrane viscosity on the pulse propagation can be neglected \cite{rubi78}. Concerning the viscosity of the axoplasm, for the general case of a viscous compressible fluid enclosed in an elastic tube, the dispersion equation is quite complicated and has not been solved in closed form. However, the equation is simplified and can be solved in the high viscosity limit, defined by:
\begin{equation}
\alpha \equiv R \sqrt{\frac{\omega\rho}{\mu}} \ll 1.
\end{equation}
Here we evaluate eq. (10) for typical axons. As before, we assume that the duration of the axoplasmic pressure pulse is similar to that of the action potential, about 1 ms, yielding $\omega \approx 3142$ rad/s. Axoplasm viscosity $\mu$, as is relevant for perturbations associated with small-amplitude axoplasmic pressure pulses, is assumed to be similar to the viscosity of water, $\mu=6.82 \cdot 10^{-4}$ Pa$\cdot$s at 38 $^\circ$C \cite{lide02}. Taking $2R=10 \mu$m (the typical internal myelin diameter for A$\alpha$ myelinated fibers) and $\rho = 1050$~kg/m$^3$ yields $\alpha = 0.35$, which should be small enough for high-viscosity approximation to the dispersion equation to hold \cite{rubi78} for axons of similar or smaller diameters. When $\alpha\gg 1$ ($2R\gg29\mu$m), the effects of viscosity are small, and the pressure pulse propagates with little decay and with the velocity described by eq. (9).

When eq. (10) holds, the phase velocity of axoplasmic pressure waves (the velocity of the harmonic waves) is given by \cite{rayl45,morg54,rubi78}:
\begin{equation}
v_{ph}=\frac{c}{2} \cdot \alpha \cdot v_0,
\end{equation}
where $\alpha$ and $v_0$ are given by eqs. (10) and (9), respectively; $c=1$ when $\frac{2R}{K} \ll \kappa$ (the limit of a rigid membrane), while $c=\frac{2}{\sqrt{5-4\nu}}$ when  $\frac{2R}{K} \gg \kappa$ (the limit of a soft membrane). The factor $c$ accounts for movement of the membrane in the axial direction caused by the viscous axoplasm  in the limit of a soft membrane; here and below we assume $\nu=0.5$ for the incompressible lipid bilayer, leading to $\frac{2}{\sqrt{5-4\nu}}\approx 1.15$.
The velocity of the pressure pulse (the "group" velocity) is given by $v_{gr}=\frac{\partial \omega}{\partial k}$,
where $k$ is the wavenumber.
From eqs. (9) - (11), and using the relations $v_{ph}=\frac{\omega}{k}$ and $v_{gr}=\frac{\partial \omega}{\partial k}$, we obtain the velocity of the axoplasmic pressure pulse in the high viscosity limit (10):
\begin{equation}
v_{gr}=c \cdot \alpha \cdot v_0 = c \cdot R \sqrt{\frac{\omega}{\mu(\kappa +\frac{2R}{K})}}.
\end{equation}
Under the same limit, the distance over which the pulse amplitude decreases $e$-fold (the decay length), is \cite{rayl45,morg54,rubi78}:
\begin{equation}
l = \frac{c}{2} \cdot R \sqrt{\frac{1}{\omega \mu (\kappa +\frac{2R}{K})}}.
\end{equation}

\subsection{Myelinated axons}
In the Hodgkin-Huxley theory of propagation of the action potential, the role of the myelin sheath is in insulating the membrane electrically and in decreasing the membrane electrical capacitance; both contribute to more rapid conduction of the action potential. However, the myelin sheath is also quite rigid. About 80\% of dry myelin content is lipid, with a roughly 2:2:1 molar ratio of the three major lipids, cholesterol, phospholipids and galactolipids \cite{knaa05}. Cholesterol is known to increase the strength and elastic modulus of lipid bilayers, with the highest modulus and highest strength achieved for 50 mol\% cholesterol \cite{need95}, i.e., rather close to the concentration in myelin. The remaining 20\% of dry myelin content consists of protein, including various types of fibrous proteins that can form rigid structures. Altogether, for the deformation of circumferential strain, the myelin elastic area expansion modulus is about 2 N/m for a single myelin layer of 4 nm thickness \cite{need95}. Noting that $K =Eh$ (for $\nu=0.5$), the myelin effective Young's modulus is $E = 5\cdot10^8$ Pa. The ratio of the axon diameter (internal myelin diameter) to the nerve fiber diameter (external myelin  diameter) is typically about 0.7 \cite{dona05}, implying a myelin thickness of $h\approx0.43 R$. Therefore, $\frac{2R}{K}=\frac{2R}{Eh} \approx 9.3\cdot 10^{-9}$ Pa$^{-1}$. This is about 23 times greater than the compressibility of water at 38$^\circ$C, $\kappa =4.04\cdot10^{-10}$ Pa$^{-1}$. Hence, $v_0$, as used in eq. (12), is only $\sqrt{24}\approx4.9$ times less than the speed of sound in water (see eq. (9)). In other words, the myelin sheath is so rigid that its distensibility decreases the pulse speed only about 4.9 times compared to the theoretical maximum if axon walls were absolutely rigid. 
For typical myelinated axons, e.g., cat myelinated axons of 10 $\mu$m fiber diameter at 38 $^\circ$C, eq. (12) yields $v_{gr} = 88$~m/s (using $E=5\cdot10^8$ Pa, $2R=7 \mu$m, $h=0.43 R$, $\nu=0.5$, $\omega=3142$ rad/s, $\mu=6.82 \cdot 10^{-4}$~Pa$\cdot$s and $\kappa =4.04\cdot10^{-10}$~Pa$^{-1}$). Given the roughness of the estimate, this theoretical value of 88 m/s is in excellent agreement with the measured action potential velocity of 60 m/s for these fibers \cite{hurs39}.
From eq. (13), the decay length for the propagation of the pulses in myelinated segments of such a fiber (where the pulse propagates passively) is 14 mm. Assuming that the distance between the nodes of Ranvier is 1 mm (100 times the external diameter of myelin), the pulse amplitude decays 7\% over the internodal distance.

\subsection{Unmyelinated axons}
Unmyelinated nerve fibers usually have diameters of 0.1 - 1.2 $\mu$m. The action potential conduction velocity, in m/s, is given approximately by $v\approx 1800\sqrt{R}$, where $R$ is the axon radius in meters \cite{hobb07} (values quoted in the literature range from $v\approx 1000\sqrt{R}$ \cite{plon88} to $v\approx 3000\sqrt{R}$ \cite{rush51}).
Therefore, for an axon of 1~$\mu$m diameter, conduction velocity is about 1.27~m/s.
Assuming $2R=1$ $\mu$m, $K=0.8$ N/m (twice the single plasma membrane value of 0.4 N/m \cite{thom97}), and $\omega$, $\mu$, $\kappa$ and $\nu$ as before for myelinated axons, eq. (12) yields a pressure pulse velocity of $v_{gr} = 1.11$ m/s, very close to the experimental value. From eq. (13), the decay length for this fiber is 0.18 mm. It is truly remarkable that, for both myelinated and unmyelinated fibers, the predicted velocity of propagation of axoplasmic pressure pulses is very close to the measured velocities of nerve impulses. 

\subsection{Dependence of propagation velocity on fiber diameter}
As can be seen from the above numerical estimates, for both myelinated and unmyelinated axons the condition $\frac{2R}{K} \gg \kappa$ is true; and therefore, eq. (12) simplifies to:
\begin{equation}
v_{gr}= \sqrt{\frac{2 R \omega K}{3 \mu}}.
\end{equation}
This relationship shows that for unmyelinated axons, the axoplasmic pressure pulse velocity is proportional to the half-power of the axon diameter, assuming that the pulse duration, the membrane area expansion modulus and the axoplasm viscosity do not vary with axon diameter. 
For myelinated axons, introducing $\gamma =h/R$ (the ratio of the myelin sheath thickness to the axon radius) and using $K=Eh$, we can rewrite eq. (14) as:
\begin{equation}
v_{gr}= R\sqrt{\frac{2\omega E \gamma}{3 \mu}}.
\end{equation}
Therefore, for myelinated axons, assuming that pulse duration and specific properties of axoplasm and myelin are independent of diameter and that the myelin sheath thickness scales in proportion to the fiber diameter (i.e., $\gamma$ is constant), the pressure pulse velocity scales linearly with the fiber diameter. It is interesting that a linear dependence of pressure pulse velocity on fiber diameter is also obtained for myelinated axons without assuming $\kappa\ll\frac{2R}{K}$, from eq. (12):
\begin{equation}
v_{gr}= c \cdot R \sqrt{\frac{\omega}{\mu(\kappa +\frac{2}{E \gamma})}}.
\end{equation}
In summary, given that specific properties of axoplasm and myelin are independent of fiber diameter and that the myelin sheath thickness scales in proportion to fiber diameter, the axoplasmic pressure pulse velocity scales as $\sqrt{R}$ for unmyelinated axons and as $R$ for myelinated axons, similar to what is measured experimentally \cite{hobb07,rush51}. 

\subsection{Dependence of propagation velocity on temperature}
The duration of the action potential increases with decreasing temperature, e.g., with Q$_{10}$ (the ratio of the velocity at one temperature to the value at a temperature 10 $^\circ$C lower) of 3.4 at 37 $^\circ$C for cat vagus myelinated fibers \cite{paint66}. Assuming that the duration of the axoplasmic pressure pulse is similar to that of the action potential, taking the change in $\mu$ from 37 $^\circ$C to 27 $^\circ$C as for water, Q$_{10}=0.81$ \cite{lide02}, and assuming that the other parameters in eq. (12) do not change with temperature, we obtain the predicted Q$_{10}$ of 2 for the nerve impulse propagation velocity. This is rather close to the experimental value of 1.6 \cite{paint65}.

\section{Discussion}
It has recently been suggested \cite{heim05,heim06} that the Hodgkin-Huxley model \cite{hodg52} of the action potential does not provide a satisfactory description of the nerve impulse because it does not include the mechanical  \cite{iwas80b,iwas80,tasa80,tasa82,tasa90} and optical \cite{tasa68,tasa69} changes associated with the action potential, as well as it is inconsistent with observed reversible changes in temperature and heat \cite{abbo58,howa68,ritc85,tasa89,tasa92}. Furthermore, the famous Meyer-Overton rule \cite{meye99,over01,urba06} is seemingly inconsistent with the Hodgkin-Huxley theory, as it implies an action of anesthetics other than binding to ion channels \cite{heim06}. The authors suggested a theory of the action potential based on solitons propagating in the axon membrane \cite{heim05,heim06}. Although this theory is in principle able to account for the abovementioned phenomena, apparently it cannot explain changes in the propagation velocity that are related to axon diameter.

Following earlier work \cite{rva03,rva03b}, we have here proposed an alternative theory of the propagation of the nerve impulse. We suggested that the traveling transmembrane voltage spike (the action potential) is preceded by a pressure pulse propagating in the axoplasm. We have shown that, under physiological conditions, such pulses decay over roughly 1 mm distances; therefore, sustained propagation requires a mechanism of amplification. Based on known axon properties, we have proposed several amplification mechanisms. Each of these mechanisms involves the Hodgkin-Huxley voltage spike as a mediator, thereby providing the spike a functional role within the model. Due to the presence of the axoplasmic pressure pulse, the model is consistent with the mechanical, optical and thermal changes associated with the action potential  \cite{iwas80b,iwas80,tasa80,tasa82,tasa90,abbo58,howa68,ritc85,tasa89,tasa92,tasa68,tasa69}. Importantly, the simple and powerful Meyer-Overton rule \cite{meye99,over01,urba06}, valid over 5 orders of magnitude, currently appears to be difficult, if not impossible, to interpret within the Hodgkin-Huxley theory. Within the presented framework, however, the Meyer-Overton rule is readily explained based on properties of lipid bilayers. A simple test of the presented theory is its ability to predict the velocity of the nerve impulse. We have shown that, over 2 orders of magnitude (for myelinated and unmyelinated axons), the predicted velocity is close to that which is experimentally observed, despite the fact that the precision of calculations is limited by current knowledge of the mechanical properties of axons and of the duration of the axoplasmic pressure pulse. The dependence of nerve impulse velocity on fiber diameter and temperature is also reproduced well by the model. A transmembrane voltage spike such as the action potential, however, is not a necessary feature of the model; therefore, it is suggested that some neurons may not exhibit the voltage spike, while still transmitting the nerve signal over short distances. 

\begin{acknowledgments}
The author gratefully thanks his father, Prof. Michael A. Rvachov, for 
useful discussions.
\end{acknowledgments}

\bibliographystyle{elsart-num}
\bibliography{proj}

\end{document}